\documentclass[usenatbib]{mnras}

\usepackage[T1]{fontenc}
\usepackage{ae,aecompl}

\usepackage{hyperref} 
\usepackage{longtable} 
\usepackage{threeparttablex}  
\usepackage{array}  

\usepackage{graphicx}	
\usepackage{amsmath}	
\usepackage{amssymb}	

\newcommand{\bc}{\begin{center}}
\newcommand{\ec}{\end{center}}

\newcommand{\Msun}           {\,{\rm M}_\odot}

\title[Dwarf Galaxy SFHs] {The star formation histories of dwarf galaxies in Local Group cosmological simulations} 

\author[R. Digby et al.]{
\parbox[t]{\textwidth}{Ruth Digby$^{1}$\thanks{Email: digbyr@uvic.ca}, Julio F. Navarro$^{1,2}$,
                       Azadeh Fattahi$^{3}$, Christine M. Simpson$^{4,5}$, Kyle A. Oman$^{6}$, 
                       Facundo A. Gomez$^7,8$, Carlos S. Frenk$^{3}$, Robert J. J. Grand$^{9}$, 
                       Ruediger Pakmor$^{9}$} \\ \\
\parbox[t]{\textwidth}{
  $^1$Department of Physics and Astronomy, University of Victoria, 
      PO Box 3055 STN CSC, Victoria, BC, V8W 3P6, Canada\\
  $^2$Senior CIfAR Fellow.\\
  $^3$Institute for Computational Cosmology, Department of Physics, 
      University of Durham, South Road, Durham DH1 3LE, UK\\
  $^4$Enrico Fermi Institute, The University of Chicago, Chicago, IL 60637, USA \\
  $^5$Department of Astronomy \& Astrophysics, University of Chicago, Chicago, IL 60637, USA \\ 
  $^6$Kapteyn Astronomical Institute, University of Groningen, Postbus 800, 
      NL-9700 AV Groningen, the Netherlands\\ 
  $^7$Instituto de Investigaci\'on Multidisciplinar en Ciencia y Tecnolog\'ia, Universidad 
      de La Serena, Ra\'ul Bitr\'an 1305, La Serena, Chile\\
  $^8$Departamento de F\'isica y Astronom\'ia, Universidad de La Serena, Av. Juan Cisternas 
      1200 Norte, La Serena, Chile\\
  $^9$Heidelberger Institut fur Theoretische Studien Heidelberg, Baden-Württemberg, DE\\
  }
}

\date{Accepted XXX. Received YYY; in original form ZZZ}

\pubyear{2018}

\begin{document}
\label{firstpage}
\pagerange{\pageref{firstpage}--\pageref{lastpage}}
\maketitle

\begin{abstract}
We use the APOSTLE and Auriga cosmological simulations to study the star formation histories (SFHs) of field and satellite dwarf galaxies. Despite sizeable galaxy-to-galaxy scatter, the SFHs of APOSTLE and Auriga dwarfs exhibit robust average trends with galaxy stellar mass: faint field dwarfs ($10^5<M_{\rm star}/M_\odot<10^{6.5}$) have, on average, steadily declining SFHs, whereas brighter dwarfs ($10^{7.5}<M_{\rm star}/M_\odot<10^{9}$) show the opposite trend. Intermediate-mass dwarfs have roughly constant SFHs. Satellites exhibit similar average trends, but with substantially suppressed star formation in the most recent $\sim 5$ Gyr, likely as a result of gas loss due to tidal and ram-pressure stripping after entering the haloes of their primaries. These simple mass and environmental trends are in good agreement with the derived SFHs of Local Group (LG) dwarfs whose photometry reaches the oldest main sequence turnoff. SFHs of galaxies with less deep data show deviations from these trends, but this may be explained, at least in part, by the large galaxy-to-galaxy scatter, the limited sample size, and the large uncertainties of the inferred SFHs. Confirming the predicted mass and environmental trends will require deeper photometric data than currently available, especially for isolated dwarfs. 
\end{abstract}

\begin{keywords}
Local Group -- galaxies: dwarf -- galaxies: star formation -- galaxies: evolution 
\end{keywords}

\section{Introduction}
\label{sec_intro}

Understanding dwarf galaxies is integral to a comprehensive picture of galaxy evolution. In the hierarchical model of galaxy formation, today's massive galaxies were formed through the successive merging of smaller objects so, in a sense, every galaxy, however massive, was once a dwarf \citep{White1991_HierarchicalClustering}. Furthermore, dwarfs are extremely useful tools to study the many processes governing galaxy evolution  \citep{Mateo1998}. As the most numerous galaxies in the Universe, dwarfs probe a wide range of environments. Some evolve in near isolation, making them ideal targets to study internal drivers, such as gas accretion rates and energetic feedback from evolving stars. Others were accreted into the potential wells of larger systems and are affected by external effects such as tidal  \citep{Mayer2001_TidalStirring, Kravtsov2004_TidalStripping, Fattahi2018_TidalStripping} and ram-pressure \citep{Gunn1972_RPstripping, Abadi1999_RPstripping} forces. Because of the shallow potential wells of dwarfs, these perturbations often leave an imprint in their present-day structure and star formation history. 

Dwarf galaxies have traditionally been classified according to their current star formation activity into dwarf spheroidal (dSph) systems with no gas and, consequently, no ongoing star formation; and into dwarf irregular (dIrr) systems where gas is presently turning into stars at appreciable rates \citep{Hodge1971}. A third category of `transition' (dT) systems is also often invoked to denote systems with recent star formation but no massive stars or HII regions \cite[see, e.g., the review by][and references therein]{Tolstoy2009}.

It has long been appreciated that, however practical from a morphological standpoint, this categorization provides limited physical insight, as it is heavily weighted by the present-day state of a system, which may be transient and, generally, a poor proxy for the evolutionary history of a dwarf. Indeed, some dSph and dIrr systems share many structural properties and evolutionary characteristics, and differ only because in the latter star formation continues to this day, whereas it has ceased (often quite recently) in the former \citep{Grebel1999, Tolstoy2009, Weisz2011_SFHof60DGs, Gallart2015_FastSlow}.

A more comprehensive view is provided by the star formation history (SFH) of a dwarf, which describes the mass-weighted distribution of the formation times of its long-lived stars. SFHs can be estimated from deep color-magnitude diagrams (CMDs) of their resolved stellar population, a field of study that has been largely enabled by the advent of panoramic imaging capabilities at the Hubble Space Telescope and by the development of sophisticated modelling algorithms that reliably synthesize the various stages of stellar evolution \citep[see, e.g.,][]{Dolphin2002_MATCH, Hidalgo2011_IACroutine, Weisz2011_SFHof60DGs}.

There are now estimated SFHs for $\sim 100$ dwarf galaxies in our local Universe (some as far away as $\sim 5$ Mpc), spanning a wide range of stellar masses, morphological types, and environments \citep{Weisz2011_SFHof60DGs, Weisz2014_SFHofLGDGs, Gallart2015_FastSlow, Skillman2017_SFHislands}. These SFHs have enabled a quantitative characterization of the vast morphological diversity of dwarf galaxies, and have provided important clues to the main mechanisms governing their evolution.

The measured SFHs have also elicited questions that so far have not been properly answered. One of them is the role of the environment. Satellites of the Milky Way (MW) and M31 do not currently form stars, unless they are quite massive (such as the Magellanic Clouds). Nearly all isolated (`field') dwarfs, on the other hand, are star-forming \citep{Geha2012}, except for a few puzzling cases, like the Cetus and Tucana dSphs \citep{Monelli2010_Cetus, Monelli2010_Tucana}. These exceptions indicate that environment plays a nuanced role in regulating star formation that is still not fully understood.

A second issue concerns the earliest and latest stages of star forming activity. All satellites apparently started forming stars very early on, but differ widely on when star formation ceased. Available data show no obviously discernible dependence on distance to the host, and suggest a puzzling distinction between M31 and MW dSphs: Carina, Fornax, and Leo I stopped forming stars only $2$--$3$ Gyr ago but no known M31 satellite ceased forming stars so late \citep[and references therein]{Weisz2014_MWvsM31sats}

The role of cosmic reionization is also unclear. Although eminently necessary on theoretical grounds to curtail the fraction of baryons able to form stars in low-mass systems \citep{Efstathiou1992, Bullock2000, Benson2002, RicottiGnedin2005}, there are apparently no `smoking gun' signatures left by this process that can be read directly from the SFHs \citep{GrebelGallagher2004, Okamoto2009}.

In addition, SFHs show no obvious dependence on the stellar mass of the dwarf;  is this because trends are weak and easily masked by large galaxy-to-galaxy scatter and the still relatively small number of systems surveyed, or a result of deeper physical significance?

Finally, the sheer diversity of SFHs is a puzzle in itself: what drives galaxies with similar stellar masses, presumably inhabiting similar mass haloes, and in similar environments, to exhibit the bewildering array of evolutionary histories their CMDs suggest?

We analyse these issues here using the star formation histories of simulated dwarf galaxies in regions of the Universe selected to resemble the Local Group. The simulations are mainly taken from the APOSTLE\footnote{A Project Of Simulating The Local Environment} project \citep{Fattahi2016_APOSTLEintro, Sawala2016_APOSTLEintro}, a suite of LCDM cosmological hydrodynamical simulations which follow a volume that matches fairly well that where the $\sim 100$ dwarfs with observed SFHs are located. We also use results from an independent simulation project \citep[Auriga;][]{Grand2017_AurigaIntro} to assess the robustness of our results and their reliability to different simulation methodology.

This paper is organized as follows. Sec.~\ref{SecSims} introduces the simulations, describes the simulated galaxy sample, and explains the procedure to estimate SFHs. In Sec.~\ref{SecObs} we describe the available observational data for the Local Group. Sec.~\ref{SecSimResults} presents our main findings for simulated satellites and field dwarfs. Sec.~\ref{SecApoObs} compares our main findings with observed LG trends. We conclude with a brief summary of our conclusions in Sec.~\ref{SecConcl}.

\section{The APOSTLE and Auriga Simulations}
\label{SecSims}

We describe below the APOSTLE and Auriga cosmological hydrodynamical simulations used in our analysis, as well as the galaxy sample selection procedure and the methods adopted to study SFHs.

\subsection{APOSTLE}
\label{SubSecApostle}

APOSTLE consists of a suite of 12 Local Group-like volumes, selected from a $\Lambda$CDM N-body cosmological simulation of a $100^3$ Mpc$^3$ periodic box \citep[DOVE;][]{Jenkins2013_DOVE}. Volumes were selected to reproduce the kinematic properties of the MW-M31 pair and their surrounding environment out to $\sim 3$ Mpc. Each volume was resimulated using the `zoom-in' technique \citep[e.g.,][]{Frenk1996, Power2003_ZoomIn}, at three different numerical resolutions (L1, L2, L3, with gas particle masses of $\sim10^4, 10^5, 10^6 \Msun$, and gravitational Plummer-equivalent softening lengths of $134$, $307$, and $711$ pc, respectively).  All APOSTLE volumes have been simulated at level L2 and L3, but to date only five volumes have been run at the highest resolution (Ap-L1).  We restrict our analysis here to the Ap-L1 and Ap-L2 realizations of these five volumes.

The simulations were performed using a modified version of the TreePM-SPH code P-Gadget3 \citep{Springel2008b_PGadget3}, developed for the EAGLE project \citep{Schaye2015_EAGLEintro,Crain2015_EAGLEintro}. The subgrid galaxy formation model of EAGLE includes photoionization due to an X-ray/UV background\footnote{Hydrogen ionization happens instantaneously at z=11.5}, metallicity-dependent gas cooling and star formation, stellar evolution and supernova feedback, black-hole accretion and AGN feedback  (although $<1\%$ of our $z=0$ dwarfs contain black holes that have grown beyond the seed mass); and it was calibrated to approximately match the average size of the stellar component of galaxies and to reproduce the $z=0.1$ stellar mass function of galaxies down to $M_{\rm star}\sim 10^8\, \Msun$. We refer the interested reader to \citet{Schaye2015_EAGLEintro} and references therein for full details. The APOSTLE simulations show that the same subgrid physics can reproduce the stellar mass function of satellites in the Local Group down to $M_{\rm star}\sim 10^5 \Msun$, without further recalibration \citep{Sawala2016_APOSTLEintro}.

\subsubsection{APOSTLE galaxy sample}
\label{SubSubSecApoGals}

Dark matter haloes in APOSTLE are identified using the friends-of-friends (FoF) algorithm \citep{Davis1985_FoF} with linking length of 0.2 times the mean interparticle separation. Gas and star particles are assigned to the FoF groups according to their nearest DM particle. Bound (sub)structures within each FoF group are then found iteratively using Subfind \citep{Springel2001a_Subfind, Dolag2004} on stars, gas, and DM particles.

Galaxies are defined as the baryonic components of these subhaloes within a `galactic radius' $r_{\rm gal} = 0.15\, r_{200}$, where $r_{200}$ is the virial\footnote{Virial quantities are defined within a radius, $r_{200}$, enclosing a mean density 200 times the critical density for closure. A subscript `200' identifies quantities defined within or at that radius.} radius: $r_{\rm gal}$ is found to contain essentially all of the stars and star-forming gas in a halo. Satellite galaxies, defined as those which inhabit subhaloes other than the main (`central') object in each FoF group, do not have a well-defined virial radius. In these cases, we follow \citet{Fattahi2018_TidalStripping} and use the average relation between $r_{\rm gal}$ and the maximum circular velocity, $V_{\rm max}$, for central galaxies in APOSTLE to define $r_{\rm gal}/$kpc$ = 0.169(V_{\rm max}/$km s$^{-1})^{1.01}$.  The relation between $r_{200}$ and $V_{\rm max}$ is very tight, so using this same definition of $r_{\rm gal}$ for all galaxies (field and satellites) gives equivalent results.

We will refer to the two main galaxies in each volume as the `Milky Way and M31 analogues' or, more generally, as the `primary' galaxies of each volume. Dwarf galaxies within 300 kpc of either primary are defined as `satellites,' and more distant dwarfs as `field' galaxies, provided they are the central object of their FoF group. We restrict our analysis of field galaxies to those within $2$ Mpc of the barycentre of the primaries. Beyond  $\sim 3$ Mpc, simulated galaxies are contaminated by low-resolution boundary particles. For completeness, we include all simulated galaxies in our analysis, but recommend caution when interpreting those resolved with fewer than $10$ star particles. This corresponds to a stellar mass of $\sim 10^5 \, M_\odot$  in the case of Ap-L1 runs, and $\sim 10^6 \, M_\odot$ for Ap-L2 runs. We focus on dwarf galaxies in this study, so our sample retains only simulated dwarfs  with $M_{\rm star}<10^9\, M_\odot$.

\subsection{Auriga}
\label{SubSecAuriga}

Auriga consists of zoom-in resimulations of $\sim 30$ relatively isolated Milky Way-sized haloes (i.e., virial mass of order $\sim 10^{12}\, M_\odot$), and their surrounding volumes. To date, $6$ have been run at the highest resolution level (L3). Unlike APOSTLE, which follows regions with a pair of massive haloes separated by $\sim 1$ Mpc and on their first approach, Auriga follows individual haloes at comparable (and, in many cases, higher) numerical resolution than APOSTLE. Auriga also uses completely independent hydrodynamics and star formation/feedback subgrid modules, built on the moving-mesh code, AREPO \citep{Springel2010_AREPO}. The Auriga code is similar to that used in the Illustris Project \citep{Vogelsberger2014_IllustrisIntro}, which, like EAGLE, has been successful at reproducing the main properties of the galaxy population in cosmologically significant volumes. AREPO includes a wide array of physical processes, similar to those in the EAGLE code used for APOSTLE, although reionization is set to be complete later in Auriga (by  $z\sim 6$). As in APOSTLE, Auriga contains prescriptions for AGN feedback, but at $z=0$ none of Auriga's field dwarfs and $<1\%$ of Auriga's satellites contain black holes. We refer the interested reader to \cite{Grand2017_AurigaIntro} for details on the Auriga project.

We use here data from Auriga's L3 simulation suite. With a typical gas cell mass of $6 \times 10^3 \, M_\odot$, Au-L3 has roughly a factor of 2 higher resolution than the Ap-L1 runs. None of the $6$ volumes run at L3 have contamination within $1$ Mpc; we select dwarfs out to a distance of $800$ kpc from the primary to minimize boundary effects. We also compare results from the Au-L4 realizations of those same volumes; Au-L4 has a baryonic mass resolution of $\sim 5 \times 10^4 \, M_\odot$. As with APOSTLE, dwarfs within $300$ kpc of the primary are defined as satellites, and those between $300$ and $800$ kpc as field dwarfs, provided they are the central galaxy of their FoF group. Galaxy properties (stellar mass, etc.) are computed following a similar procedure to that described in Sec.~\ref{SubSubSecApoGals}.

\subsection{Simulated star formation histories}
\label{SecSimSFHs}

We characterize the star formation history (SFH) of each simulated dwarf by computing the fraction of stars formed in three different intervals of cosmic time, $t$: $f_{\rm old}\equiv f_{\rm o}$ refers to `old' stars ($t_{\rm form}<4$ Gyr), $f_{\rm int}\equiv f_{\rm i}$ to `intermediate-age' stars ($4<t_{\rm form}/$Gyr$<8$), and $f_{\rm young}\equiv f_{\rm y}$ to `young' stars ($t_{\rm form}>8$ Gyr). We express these fractions as star formation rates (SFR) normalized to the past average, ${\bar f}=M_{\rm star}/t_0$, where $t_0=13.7$ Gyr is the age of the Universe, and $M_{\rm star}$ is the stellar mass of a dwarf at $z=0$ \citep{BenitezLlambay2015_ImprintReionzn}. In other words, 

\begin{equation}
    f_j=\frac{1}{X}\frac{\Delta M_j/\Delta t_j}{\bar f}, 
\end{equation}

\noindent where the subscript $j$ stands for either the `old', `intermediate', or `young' component, and

\begin{equation}
    X=\frac{1}{\bar f}\sum_j \frac{\Delta M_j}{\Delta t_j}
\end{equation}

\noindent is a normalizing coefficient that ensures that $f_{\rm o}+f_{\rm i}+f_{\rm y}=1$. With this definition, galaxies that form stars at a constant rate will have $f_{\rm o} = f_{\rm i} = f_{\rm y} = 1/3$.
This procedure condenses the SFH of a galaxy of given $M_{\rm star}$ into just three numbers (two of which are independent for a galaxy of given stellar mass).

We shall also use other simple measures of the SFH that are better suited for direct comparison with observational data. These include the cumulative measures $f_{\rm XGy}$, the fractions of stars formed in the first $X$ Gyrs of evolution, as well as $\tau_{\rm X}$, defined as the time when the formation of the first X percent of the stars was completed. The simulations assume a fixed Chabrier stellar initial mass function \citep{Chabrier2003}.

\begin{figure*}
\includegraphics[page=1, width=1.\linewidth, trim = 0.cm 9.1cm 0.4cm 0cm, clip]{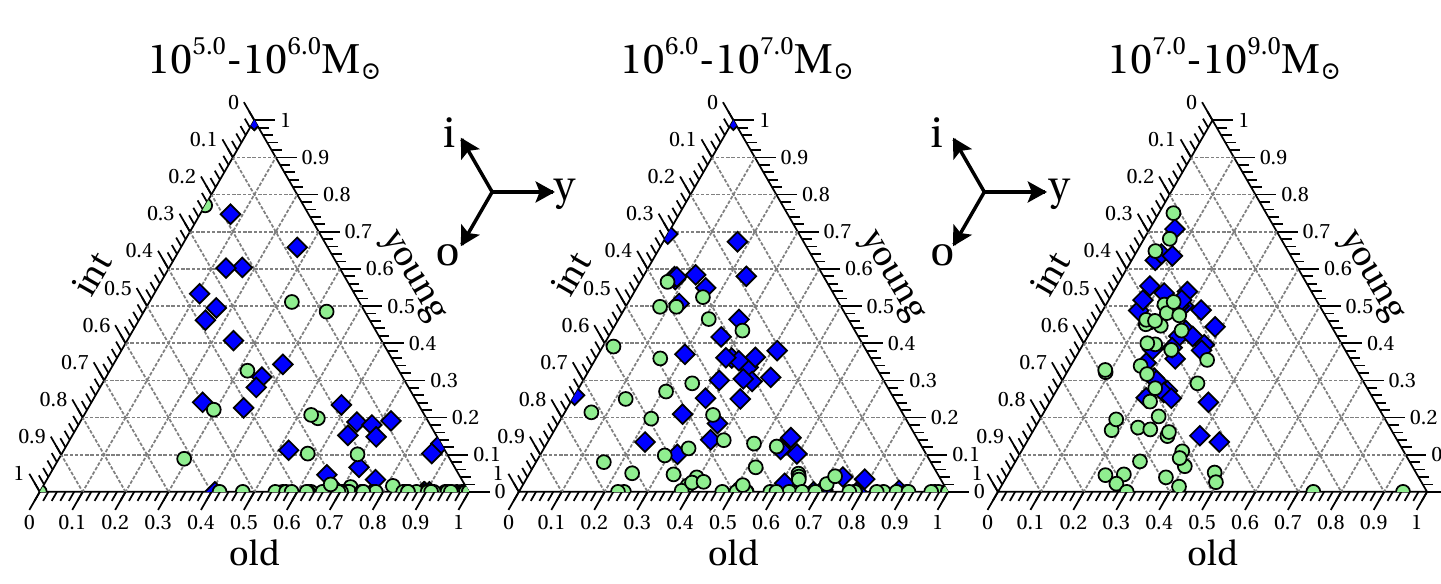}
\hspace{2em}
\includegraphics[page=1, width=1.\linewidth, trim = 0.1cm 3.6cm 0.5cm 1.4cm, clip]{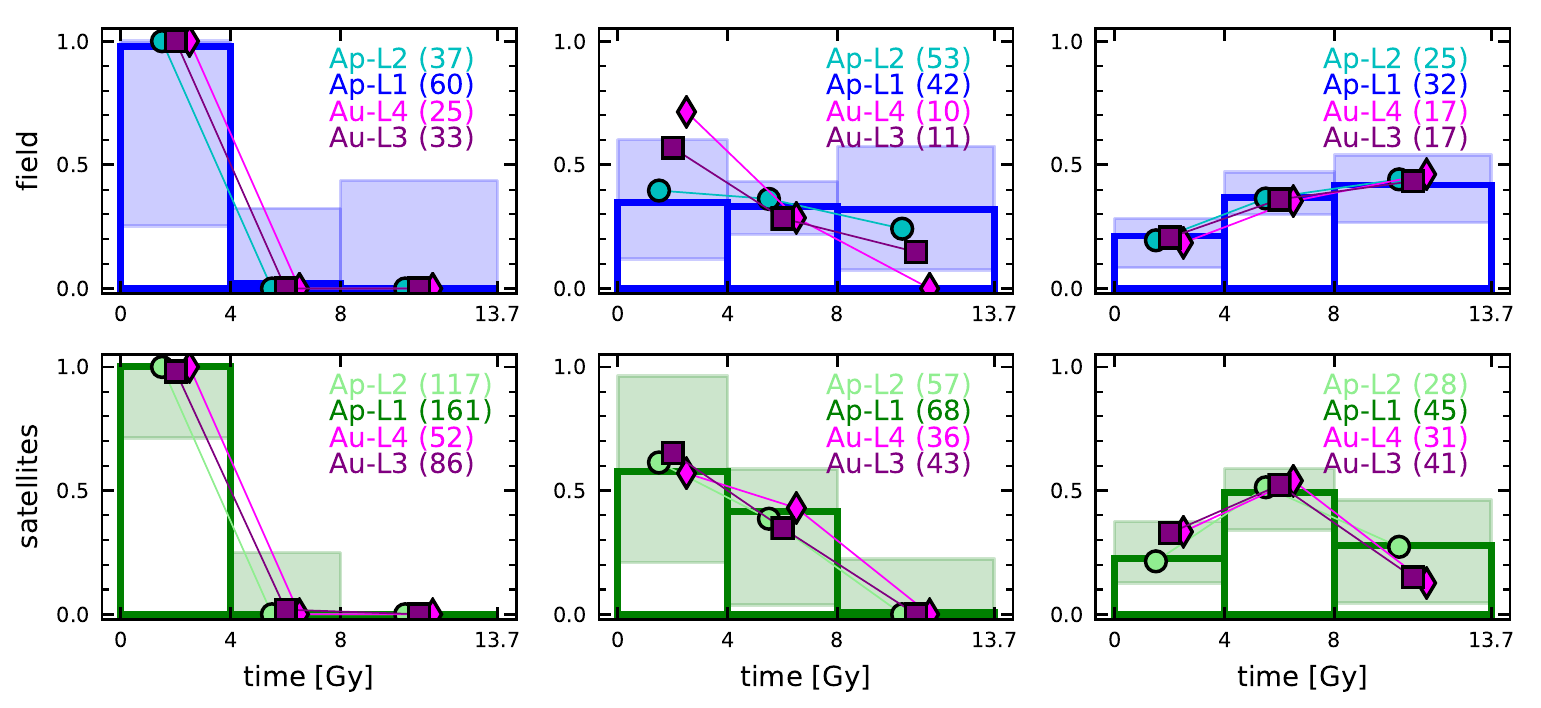}
\hspace{2em}
\caption{The star formation histories (SFHs) of APOSTLE and Auriga dwarfs. {\it Top row:} Ternary plots showing the SFHs of Ap-L1 galaxies in three bins of stellar mass, as indicated by the top legend. The arrows indicate how to read the old ($f_{\rm old}$; down and left), intermediate ($f_{\rm int}$; up and left), and young ($f_{\rm young}$; straight right) SFH fractions for each galaxy. Different symbols indicate environment: green circles correspond to satellites and the blue diamonds to field dwarfs.  {\it Middle row:} The median values of $f_{\rm old}$, $f_{\rm int}$, and $f_{\rm young}$ for field galaxies in each mass bin, as a function of cosmic time. Ap-L1 results are shown in histogram form, with shaded regions spanning the $16^{\rm th}$ to $84^{\rm th}$ percentiles. Ap-L2 results are shown with circles, Au-L3 with purple squares, and Au-L4 with magenta diamonds. For clarity, the Ap-L2 and Au-L3 markers have been offset slightly. The number of galaxies in each mass bin is given in parentheses. {\it Bottom row:} As middle row, but for satellites. Note the systematic trend with stellar mass of the average simulated field SFHs, and that said trends are robust to changes in the mass and spatial resolution of the simulations. In each mass bin, the satellite SFHs are similar to those of the isolated field galaxies, except for a significant reduction in the young stellar population.}\label{FigTernary}
\end{figure*}

\section{Local Group Observations}
\label{SecObs}

\subsection{Galaxy sample}

We will compare the simulated SFHs with available data for dwarf galaxies in the Local Group. More specifically, we use for the latter the compilations of \citet{Weisz2011_SFHof60DGs}, \citet{Weisz2014_SFHofLGDGs}, \citet{Cole2014_Aquarius}, \citet{Gallart2015_FastSlow}, and \citet{Skillman2017_SFHislands}, which provide SFHs derived from HST multi-band imaging, reduced and analysed with similar methodology. The compilation includes a total of $101$ galaxies with stellar masses in the range $6.5\times 10^{3}<M_{\rm star}/M_\odot<3.4\times 10^{9}$, $29$ of which we classify as satellites of either the MW or M31, and $72$ of which are classified as field dwarfs. The classification is based solely on distance to the nearest host; i.e., we define as satellites those within $300$ kpc of either the MW or M31, and as field dwarfs all others.

Distances and stellar masses are taken from the catalogue of \citet{Karachentsev2013_UpdatedNGC}, assuming, for simplicity, a uniform B-band mass-to-light ratio of 1 in solar units. Tables~\ref{TabField} and~\ref{TabSats} list all the galaxies selected from these compilations, together with the derived data we use in this analysis. The sample includes examples of a wide range of morphological types, including dSphs, dIrrs, and dTs, as well as the rare dwarf elliptical M32 \citep{Monachesi2012}. Not included are the Small and Large Magellanic Clouds, as their large size makes them unsuitable for study with HST's small field of view \citep{Weisz2014_SFHofLGDGs}. The farthest galaxy in the sample is $\sim 4.6$ Mpc from the Milky Way.  Note that the observed sample extends to stellar masses a bit below the $\sim 10^5\, M_\odot$ minimum mass we can resolve in the simulations. The observed sample also includes a few galaxies with $M_{\rm star}>10^9 \, M_\odot$. However, only $10$ galaxies in total are beyond the stellar mass limits of the simulated sample, so this slight mismatch is unlikely to affect adversely the conclusions of our comparison.

\subsection{Star formation histories}

Inferring SFHs from CMDs of a resolved stellar population is a mature field of study that incorporates our best understanding of the various stages of stellar evolution \citep[see][and references therein]{Dolphin2002_MATCH, Hidalgo2011_IACroutine}.  Despite these advances, SFHs derived from modelling photometric observations are still subject to substantial uncertainty, not only because of observational photometric limitations, but also because they rely on a number of assumptions such as the initial mass function, binary and blue straggler fractions, age-metallicity degeneracies, etc., which are poorly understood and difficult to account for \citep{Gallart2005_AdequacySEMsByStage}.

The modelling is also subject to substantial uncertainty in the case of observations that are not deep enough to reach confidently the oldest main sequence turnoff magnitude \citep[oMSTO; see, e.g.,][]{Gallart2005_AdequacySEMsByStage, Weisz2011_SFHof60DGs}. We shall distinguish galaxies with resolved oMSTO because there is broad agreement that models are least susceptible to systematic biases in such cases. These `oMSTO galaxies' make up about $\sim 62\%$ of our satellites and $11\%$ of our field dwarf sample. Finally, since evolving stars transit different regions of the CMD at various speeds, SFHs derived from modelling observations constrain best the cumulative fraction of stars formed by a certain time (i.e., cumulative SFHs), rather than the star formation `rate' at various times in the past.

Here we shall take the SFHs and their uncertainties directly from the references above (see also Tables~\ref{TabField} and ~\ref{TabSats}). Note that many of these SFHs are derived from fields that image only a relatively small region of the galaxy, which, in the presence of strong gradients, may bias the results. We shall neglect this complication in our comparison with simulations, and assume that the published SFHs are representative of the whole galaxy. We refer the interested reader to \citet{Gallart2005_AdequacySEMsByStage} and \citet{Weisz2014_SFHofLGDGs} for a more thorough discussion of these issues.

\section{Simulation Results}
\label{SecSimResults}

The top panels of Fig.~\ref{FigTernary} show the SFHs of Ap-L1 galaxies, split into three stellar mass bins, as indicated in the legend. Symbols of different colors are used for field dwarfs (blue diamonds) and satellites (green circles). These ternary plots  provide a convenient and economic visualization of three  parameters which add up to unity, as is the case for $f_{\rm o}$, $f_{\rm i}$, and $f_{\rm y}$. Arrows between diagrams indicate how to read each quantity along the three different axes. Galaxies that are predominantly old ($f_{\rm o}>0.5$) are found in the lower right corner, those that are predominantly young ($f_{\rm y}>0.5$) are located near the top, and galaxies where star formation peaks at intermediate epochs ($f_{\rm i}>0.5$) are found in the lower left corner. Galaxies that form stars at a near-constant average rate lie close to the centre of the plot.

The first thing to note from the top panels of Fig.~\ref{FigTernary} is the large scatter within each mass bin, for both field dwarfs and satellites. This is indicative of a strong diversity in SFHs, even for galaxies of similar mass and environment. Another notable point is that, despite the large scatter, a clear mean trend is seen with increasing stellar mass. More massive galaxies occupy the upper-left of the diagram, with lower mass systems spreading systematically  to the lower-right. In simple terms, this implies that younger stellar populations become, on average, increasingly important with increasing galaxy mass.

This is confirmed by the middle panels, which indicate, in histogram form, the median $f_{\rm o}$, $f_{\rm i}$, and $f_{\rm y}$ of the field dwarfs in the panels immediately above, along with those from Ap-L2, Au-L3, and Au-L4. The star formation rates in the low-mass bin are on average steadily decreasing, whereas those in the upper mass bin are steadily increasing. Intermediate-mass dwarfs have average SFHs closer to constant in time.

Satellites show similar trends to those of field dwarfs, including the large galaxy-to-galaxy scatter in each mass bin. The average satellite SFHs are summarized in the bottom panels of Fig.~\ref{FigTernary} and show that they are not dissimilar to those of field dwarfs, except for less prominent young stellar components. Indeed, many satellites are very close to the bottom axis of the ternary plots, which denote $f_{\rm y}=0$. In addition, the ratio of old to intermediate populations is quite similar in both satellites and field dwarfs. To first order, then, and in terms of their SFHs, satellite galaxies evolve just like regular field dwarfs, except for a substantial reduction in their ability to form stars in recent times. These trends are unlikely to be impacted by stellar stripping: most satellite dwarfs have only lost a modest fraction of their mass to tides. Fewer than $10\%$ have lost more than half of their initial infall stellar mass to tides. 

Interestingly, the trends described above are quite robust to changes in numerical resolution and simulation method. The circles in the middle and bottom panels of Fig.~\ref{FigTernary} indicate the median $f_{\rm o}$, $f_{\rm i}$, and $f_{\rm y}$ of similar samples of simulated galaxies drawn from the Ap-L2 realizations of the same volumes. Squares and diamonds show the median SFH of simulated Au-L3 and Au-L4 galaxies, respectively. Despite the order of magnitude difference in mass resolution between Ap-L1 and Ap-L2 (the gas particle mass in Ap-L2 runs is $\sim 10^5 \, M_\odot$ compared to  $\sim 10^4 \, M_\odot$ in Ap-L1), and the differences in the hydrodynamical treatment and subgrid physics between APOSTLE and Auriga, the average trends with stellar mass for all these runs in excellent agreement. This is reassuring, and suggests that the stellar mass trends discussed above are not simply the result of inadequate resolution or the choice of a particular star formation/feedback recipe. 

\subsection{APOSTLE vs Auriga}
\label{SecApAu}

\begin{figure*} 
\includegraphics[page=2, width=1\linewidth, trim = 0.5cm 3.5cm 2.1cm 0.8cm, clip]{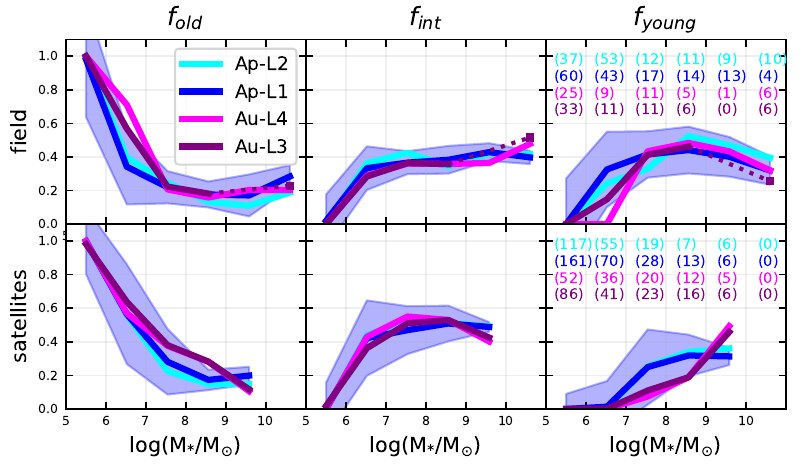} 
\caption{Median $f_{\rm old}$, $f_{\rm int} $, and $f_{\rm young}$ as a function of $M_{*}$ for Ap-L1 (high-res) and Ap-L2 (medium-res) galaxies, as well as for galaxies in Auriga ($6$ volumes at resolution level L3). The Auriga L3 suite has a nominal resolution comparable to Ap-L1. Shaded regions show 1$\sigma$ dispersion for Ap-L1 data. {\it Top.} Centrals (field dwarfs and primary galaxies). {\it Bottom.} Satellites. Numbers in parentheses indicate the number of systems in each mass bin. Note that the results for Auriga and APOSTLE are nearly identical, despite the fact that the two simulation suites use different hydrodynamical codes and independent star formation and feedback algorithms.} 
\label{FigSFHComp}
\end{figure*}

We compare APOSTLE and Auriga SFHs directly in Fig.~\ref{FigSFHComp}. This figure shows, as a function of stellar mass, the median values of $f_{\rm o}$, $f_{\rm i}$, and $f_{\rm y}$ for Ap-L1, Ap-L2, Au-L3, and Au-L4 runs. The coloured bands around the Ap-L1 results indicate the rms dispersion about the median, and is representative of the galaxy-to-galaxy variation in all four sets of simulations. As Fig.~\ref{FigSFHComp} shows, the main SFH trends in both APOSTLE and Auriga agree quite well, for both field and satellite galaxies. The agreement between these two sets of independent simulations again suggest that the mass and environmental trends highlighted in Fig.~\ref{FigTernary} are not simply artefacts of the APOSTLE subgrid physics implementation, but rather a robust characterization of the star formation activity in low-mass LCDM haloes.

For clarity and ease of presentation, the remainder of our analysis will show results from Ap-L1 only. The corresponding figures with Au-L3 data can be found in Appendix~\ref{appendix_auriga}.

\section{Simulated vs Observed Local Group SFH's}
\label{SecApoObs}

\begin{figure*}
\includegraphics[page=3,width=1\linewidth]{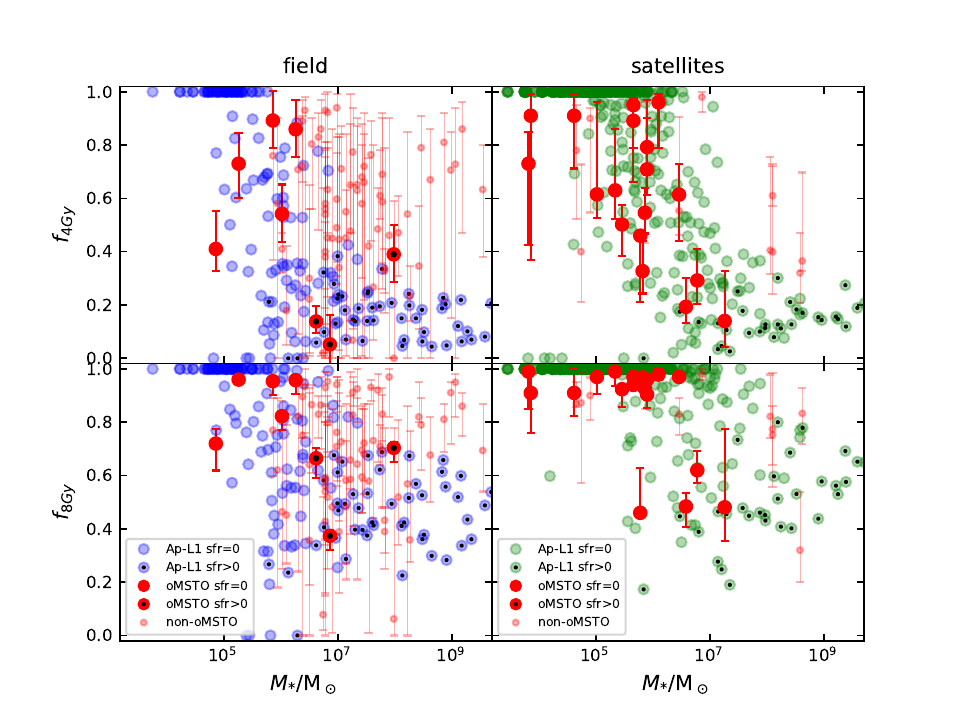}
\caption{The fraction of stars formed in the first $4$ ($f_{\rm 4Gy}$) and $8$ ($f_{\rm 8Gy}$) Gyr of cosmic evolution, as a function of stellar mass. APOSTLE galaxies are shown in blue (field dwarfs) and green (satellites); observed galaxies are in red. Error bars in the latter indicate the $16^{\rm th}$ and $84^{\rm th}$ percentile bounds on the combined statistical and systematic uncertainties, as given in the literature. SFHs published in \citet{Gallart2015_FastSlow}, which make up $6$ of the $8$ oMSTO field dwarfs, do not quote systematic uncertainties. We assign them the median error of the other oMSTO galaxies (see Tables~\ref{TabField} and ~\ref{TabSats}). Filled red circles highlight observed galaxies where the photometry reaches the oldest main sequence turnoff, and a central black `dot' indicates the oMSTO dIrrs Aquarius, IC1613, and LeoA, which are still forming stars at the present day.} \label{FigF4F8} 
\end{figure*}

\begin{figure*}
\includegraphics[page=4, width=1\linewidth, trim = 0cm 5cm 0cm 0cm, clip]{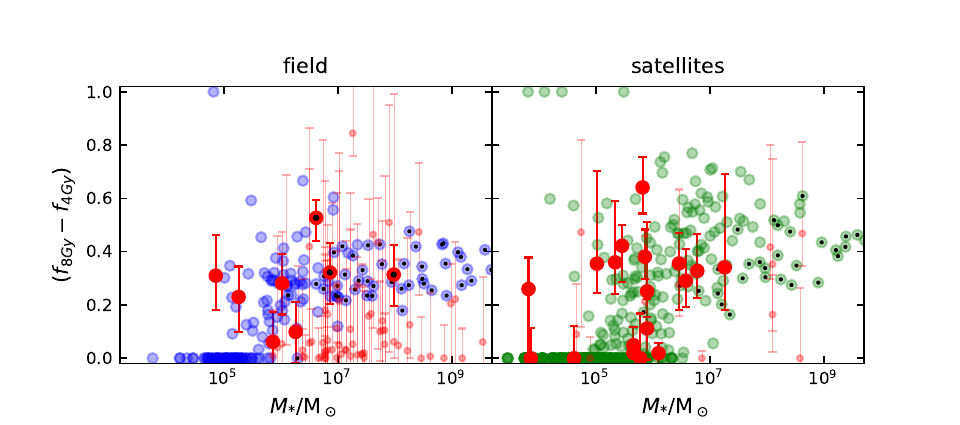}
\caption{As Fig.~\ref{FigF4F8}, but for the difference between the fraction of stars formed by the first $4$ and $8$ Gyrs of cosmic evolution. Note the large number of non-oMSTO observed galaxies (open red symbols) that appear to form no stars in that time period.} 
\label{FigF4-F8} 
\end{figure*}

\subsection{Diversity, mass and environmental trends}

We compare our APOSTLE results with observed SFHs in Fig.~\ref{FigF4F8}. We choose for this comparison two parameters that quantify the cumulative SFH of observed galaxies, namely $f_{\rm 4Gy}$ and $f_{\rm 8Gy}$ (the fraction of stars formed by cosmic time $t=4$ and $8$ Gyrs, respectively), as a function of stellar mass. These are easy to compute for simulated galaxies, and are better constrained in observations than the differential star formation `rates', especially at earlier times (i.e., large lookback times), and for observations that may lack sufficient depth to resolve the oMSTO. 

The observational data, compiled from the literature cited above, is shown in red, with error bars that span the $16^{\rm th}$ to $84^{\rm th}$ percentiles. Galaxies with resolved oMSTO are shown in solid red, others with open red circles. Ap-L1 results are shown in blue (field dwarfs) and green (satellites).

The mass trend reported above (Sec.~\ref{SecSimResults}) for APOSTLE dwarfs is also seen here: more massive galaxies have lower values of $f_{4Gy}$ and $f_{8Gy}$ than lower mass systems, indicating extended star formation activity that continues, in some cases, to the present day. (Simulated galaxies with non-zero star formation at $z=0$ are indicated with a central `dot' in the figure.) Satellites show a similar mass trend, albeit with reduced recent star formation, which translates into systematically higher values of $f_{\rm 4Gy}$ and $f_{\rm 8Gy}$ than those of field dwarfs.

Qualitatively, the same mass trends (including the substantial galaxy-to-galaxy scatter) are also followed by {\it observed} field dwarfs and satellites with photometry deep enough to reach the oMSTO (filled red circles). The mass dependence, in particular, is best appreciated in the satellite panels. Satellites, especially those of the Milky Way, make up the majority of oMSTO dwarfs because of their relative proximity. Field dwarfs are substantially farther away, and only $8$ out of $72$ have resolved oMSTO photometry. Still, the available data for those $8$ galaxies seem at face value consistent with the APOSTLE results.

The situation is less clear when considering {\it all} observed field dwarfs. Indeed, many such dwarfs have, apparently, much higher $f_{\rm 4Gy}$ and $f_{\rm 8Gy}$ values than expected from the simulations: this would imply that many field dwarfs assemble their stars much more promptly than their simulated counterparts. In addition, no obvious mass trend is seen, as opposed to the APOSTLE and Auriga results.

Before taking this discrepancy too seriously, however, one should note the very large uncertainties that apply to non-oMSTO systems (error bars indicate the $16^{\rm th}$ and $84^{\rm th}$ percentiles, and include the quoted systematic and statistical errors), which make up the majority of field dwarfs ($64$ out of $72$). These large uncertainties might not be enough to reconcile the observations with APOSTLE, however, unless there are other systematic effects at play. Indeed, most non-oMSTO field dwarfs show an intriguing feature: very similar values of $f_{\rm 4Gy}$ and $f_{\rm 8Gy}$, indicating that very few stars formed in the period $4<t/$Gyr $<8$.

This is shown in Fig.~\ref{FigF4-F8}, where it is clear that a substantial number of observed field dwarfs have $f_{\rm 8Gy}-f_{\rm 4Gy}=0$. Interestingly, none of the oMSTO field galaxies shows the same feature, and very few satellites do. Those that do have actually ceased forming stars during the first $\sim 4$ Gyr (i.e., they have $f_{\rm 4Gy}\approx f_{\rm 8Gy}\sim 1$; the same applies to most APOSTLE dwarfs that have $f_{\rm 8Gy}-f_{\rm 4Gy}=0$.) Unless there is a physical mechanism (not included in the simulations) that selectively shuts off star formation in that period, this is suggestive of some systematic effect in the SFH modelling that favours assigning old ages (i.e., $t_{\rm form}<4$ Gyr) to the majority of stars formed before $t=8$ Gyr. If this were the case, it could explain the apparent discrepancy between observations and simulations without the need to appeal for a physical mechanism that disfavours intermediate-age star formation in the field\footnote{Cosmic reionization has been invoked to explain galaxies that may have a prolonged gap in star formation activity \citep{BenitezLlambay2015_ImprintReionzn, Ledinauskas2018_ReignitedSFreionization}, but this argument is only plausible for the lowest-mass galaxies.}.

One final point to note is that of all observed satellites (right-hand panel of Fig.~\ref{FigF4F8}), the ones that deviate clearly from the APOSTLE-delineated trends are overwhelmingly non-oMSTO systems. In other words, the only satellites that clearly deviate from APOSTLE are systems where the available photometry might not be good enough to test our results. Only deeper observations of a large sample of field dwarfs will be able to clarify these issues in a conclusive manner.

\subsection{The Alpha ($A$) and Omega ($\Omega$) of star formation in dwarfs}
\label{SubSecAlphaOmega}

\begin{figure*}
\includegraphics[page=5, width=1\linewidth, trim = 0cm 5cm 1cm 0cm, clip]{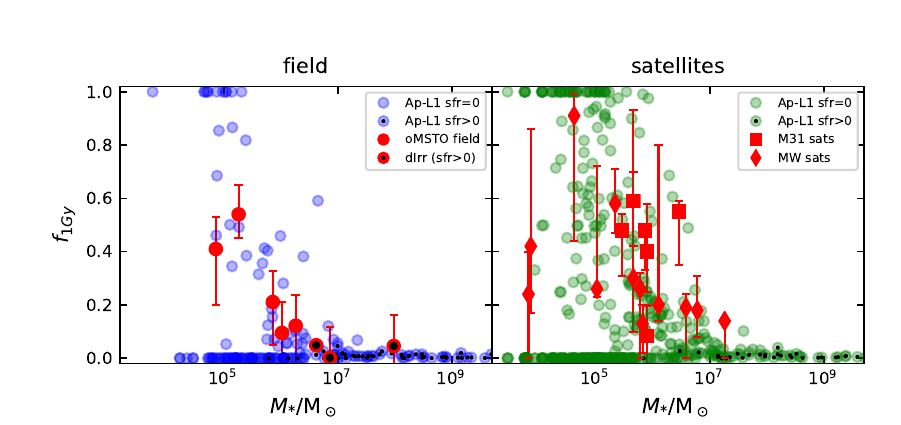} 
\caption{The cumulative fraction of stars formed in the first $\sim 1$ Gy of cosmic evolution, for APOSTLE and oMSTO galaxies only. Error bars indicate the $16^{\rm th}$ and $84^{\rm th}$ percentile bounds on the combined statistical and systematic uncertainties, as published in the literature.  Symbols differentiate observed field dwarfs (circles) from  satellites of M31 (squares) and of the Milky Way (diamonds).  Black dots indicate the dIrrs Aquarius, IC1613, and LeoA, which are still forming stars at the present day.} \label{FigMstarf1}
\end{figure*}

\begin{figure*} \includegraphics[page=6,width=1\linewidth, trim = 0cm 5cm 0cm 1cm, clip]{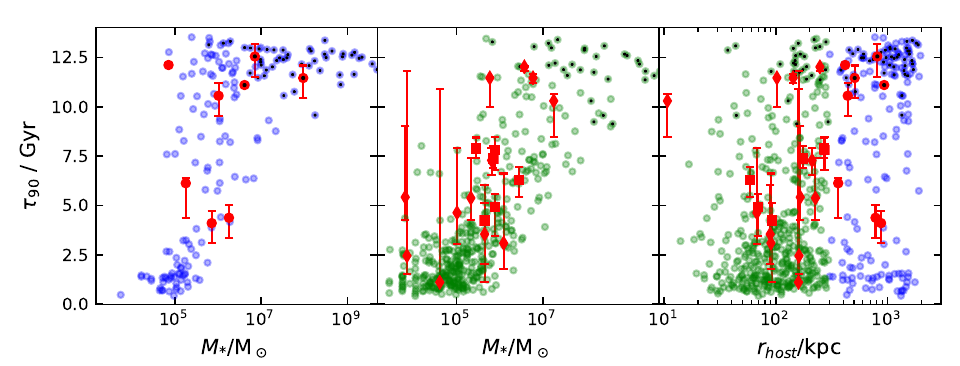} 
\caption{The cosmic time at which galaxies have formed $90\%$ of their stars, $\tau_{90}$, as a function of stellar mass (left and middle panels, showing field and satellite dwarfs, respectively) and as a function of distance from the nearest primary (right-hand panel). Values of $\tau_{90}$ are interpolated from the published SFHs.  Error bars show the corresponding width of the $16^{\rm th}$-$84^{\rm th}$ percentile error envelope given in the literature. As in Figs.~\ref{FigF4F8}-\ref{FigMstarf1}, galaxies taken from \citet{Gallart2015_FastSlow} are assigned the median error bars of all other oMSTO galaxies. Symbols differentiate observed field dwarfs (circles) from  satellites of M31 (squares) and of the Milky Way (diamonds).  Black central dots indicate the dIrrs, which are still forming stars today.} \label{FigTau90}
\end{figure*}

All dwarf galaxies appear to have a substantial population of very old stars, as if their star formation activity had started more or less synchronously at very early times. The earliest times constrained by observed SFHs correspond to roughly $t\sim 1$ Gyr (i.e., a stellar age of $\sim 12.6$ Gyr), and is only probed robustly in oMSTO galaxies. We show in Fig.~\ref{FigMstarf1} the fraction of stars formed in these galaxies  in the first $\sim 1$ Gyr of cosmic evolution, $f_{\rm 1Gy}$, and compare it with APOSTLE results.

At all masses, the majority of simulated dwarfs have very small values of $f_{\rm 1Gy}$. This is true in all environments: $\sim 70$ percent of field dwarfs and $\sim 50$ percent of satellites formed fewer than $5$ percent of their stars in the first $\sim 1$ Gyr. Observed oMSTO dwarfs, while consistent with the trend to higher $f_{\rm 1Gy}$ at low masses exhibited by some APOSTLE galaxies, lack the $f_{\rm 1Gy} \sim 0$ population that dominates the simulations.

Before reading too much into this apparent discrepancy, we caution that simulations are vulnerable to resolution effects, and sensitive to algorithmic choices, such as the star formation `thresholds' adopted, the neglect of molecular cooling, and the lack of a cold gaseous phase. The simulated results are also likely sensitive to our implementation of cosmic reionization, which is set at $z=11.5$ in APOSTLE, which corresponds to only $t\sim 0.4$ Gyr. Recent observations suggest a somewhat later reionization redshift, perhaps as low as $z_{\rm reion}\sim 5.3$, with a corresponding cosmic time of $t\sim 1.2$ Gyr \citep{Glazer2018, Planck2018}. It is therefore possible that the adoption of an early reionization redshift could have unduly reduced the fraction of stars formed in the first $\sim 1$ Gyr. Indeed, galaxies in Au-L3, which uses $z_{\rm reion}=6$, also lack the low-mass/low-$f_{\rm 1Gy}$ population and are more closely matched by observations (see Fig.~\ref{FigAuMstarf1}). On the other hand, Au-L3 satellites match observations less well; see Appendix \ref{appendix_auriga} for further discussion.

It is somewhat reassuring that the first episode of star formation in APOSTLE dwarfs occurs actually quite early in most systems. There is, however, a clear mass and resolution dependence on the age of the oldest star particle: splitting the simulated sample in the same three mass bins as in Fig.~\ref{FigTernary} ($10^{5}$-$10^{6}$; $10^{6}$-$10^{7}$; $10^{7}$-$10^{9}$, in units of $M_\odot$) we find that $90\%$ of APOSTLE dwarfs have, respectively, first-star formation times earlier than $t_A=1.2$, $0.8$ and $0.4$ Gyr for Ap-L1 runs, and $t_A=1.9$, $0.9$ and $0.5$ Gyr for Ap-L2 runs. This mass/resolution dependence shows that our estimates of $f_{\rm 1Gy}$ have not converged, and that they could easily rise in higher resolution simulations, or in simulations with a later reionization epoch. 

With this caveat, $90\%$ of all Ap-L1 dwarfs with $>10$ star particles have already started forming stars by $\sim 1.8$ Gyr, so it seems fair to conclude that essentially all simulated dwarfs do indeed have old stellar populations. This agrees qualitatively with observations, but a meaningful quantitative comparison will require simulations of much higher resolution and improved physical treatment of the formation of the first stars.

At the other extreme, Fig.~\ref{FigTau90} explores the end stages of star formation in LG dwarfs. This figure shows $\tau_{90}$ (i.e., the cosmic time when $90\%$ of star formation was completed, a robust proxy for the time when star formation effectively ceases in dSphs) as a function of stellar mass and of distance to the nearest primary. 

The agreement between simulations and observations is much better in this case. In particular, the well-defined trend of $\tau_{90}$ with stellar mass in satellites (middle panel) is well reproduced in APOSTLE. Note that if this trend were to hold at lower stellar masses it would also be consistent with the results of \citet{Brown2014}, who report that six ultra-faint dwarfs (with masses below the lower mass limit of the samples used in this paper) are consistent with having finished forming stars by $t\sim 2$ Gyr. If reionization is the culprit for the early cessation of star formation in dwarfs, then this is only clearly apparent in the faintest systems.

Simulated field dwarfs (left-hand panel in Fig.~\ref{FigTau90}) tend to fall into one of two categories: those that form stars until late times (or are still forming them at $z=0$, identified with a central `dot' in the figure), and those where star formation shuts off early on, with few examples in between. There are too few oMSTO field dwarfs for a detailed comparison, but there are no obvious deviations from this trend in the observed $\tau_{90}$. The apparent dichotomy in $\tau_{90}$ is not seen in the satellite population, where there are many systems with intermediate values of $\tau_{90}\sim 7$ Gyr. 

These trends in $\tau_{90}$ are consistent with previous studies on dwarf galaxy quiescence \citep[e.g.][]{Davies2019, Fillingham2016, Fillingham2018, Simpson2018_AurigaQuenching}, which find similar dependence on mass and environment. These authors argue that low-mass satellites may have had their star formation extinguished by the effects of ram-pressure and tidal stripping during their orbital evolution within in their host halos.


Although it is tempting to associate the secondary `peak' in the field dwarfs' $\tau_{90}$ with the claimed `synchronicity' in the cessation of star formation of some M31 and MW satellites \citep[see; e.g.,][]{Weisz2014_MWvsM31sats}, the statistical evidence seems weak, and we defer further analysis to future work. 

Galaxies marked with a central `dot' in Fig.~\ref{FigTau90} are still forming stars at $z=0$. In the case of simulations, these are overwhelmingly massive galaxies, usually with  $M_{\rm star}>10^7\, M_\odot$, in qualitative agreement with the satellites of the MW and M31, where only the most massive (e.g., the Magellanic Clouds, or M33, not included in our sample) are still forming stars today. 

Finally, the right-hand panel of Fig.~\ref{FigTau90} shows the dependence of $\tau_{90}$ on distance to the nearest host. There is no obvious dependence on distance that may be discerned from this plot, either in observed dwarfs, or in simulated ones. Our overall conclusion is that the last stages of star formation of observed galaxies are in reasonable agreement with the results of the APOSTLE simulations.

\section{Summary and Conclusions}
\label{SecConcl}

We have examined the star formation histories (SFHs) of simulated dwarf galaxies in the Local Group cosmological hydrodynamical simulations of the APOSTLE and Auriga projects. We distinguish galaxies in two environments: satellites of the primary galaxies (i.e., of the MW and M31 analogues), as well as isolated field dwarfs. Our main results may be summarized as follows.

The  SFHs of simulated dwarfs show large scatter from galaxy to galaxy, even at fixed stellar mass and similar environment. Despite the large dispersion, clear trends as a function of mass emerge when averaging over a large ensemble.

Concerning {\it field dwarfs}, the lowest mass systems we can resolve ($10^5<M_{\rm star}/M_\odot < 10^{6.5}$) have declining star formation rates: they form a large fraction of stars at early times but their star forming activity declines sharply at intermediate and recent times. Massive dwarfs ($10^{7.5}<M_{\rm star}/M_\odot < 10^{9}$) show the opposite trend: their star formation rates ramp up with time and peak at recent times. Intermediate mass dwarfs form stars at roughly constant rate, on average.

The SFHs of {\it satellite galaxies} resemble those of field dwarfs of similar mass, except for a pronounced decline in recent star formation activity. These results are insensitive to mass resolution in the APOSTLE simulations, and, encouragingly, are well reproduced in the Auriga simulation suite, which uses an independent implementation of hydrodynamics and star formation.

The comparison of these trends with those of SFHs inferred for Local Group dwarfs yields mixed but promising results. The large galaxy-to-galaxy dispersion in observed SFHs seems quite naturally reproduced by the simulations. In addition, satellites, for which much of the deepest photometry (and hence the best SFH estimates) is available, show an average mass trend also consistent with the simulation results.

The agreement between simulations and observations is more tentative for field dwarfs. Systems whose photometry reaches the oldest main sequence turnoff are, like satellites, in good agreement with APOSTLE and Auriga, but the numbers are small. Field dwarfs with shallower data (the majority) deviate systematically from the simulation predictions. In particular, there is a substantial number of systems with a prolonged `gap' in their SFH at intermediate times ($4$-$8$ Gyrs) that have no counterparts in the simulated sample. It is unclear whether this disagreement signals a failure of the dwarf galaxy formation model explored in these simulations, or systematic effects in SFHs inferred from shallow photometric data.

Assuming that the tension is resolved in favour of the dwarf galaxy formation model adopted in APOSTLE or Auriga, then the simulations would offer important insight into the physical mechanisms responsible for the results we report here.

For example, what drives the average mass trends shown in Fig.~\ref{FigTernary}? Is it differences in the fraction of retained gas after reionization, in the mass assembly history, or in the effectiveness of feedback in systems with different potential well depths?

What drives the large scatter in the SFH of dwarf galaxies at fixed mass/environment? Is is intermittent gas accretion, feedback-driven episodic star formation, interactions with the cosmic web or other external factors?

Why and when do satellites stop forming stars in recent times? Is it because of ram-pressure or tidal stripping of their extended gas envelopes? Or because star formation is enhanced, and gas consumed more quickly, in the tidal field of the host?

And finally, how can we devise tests of the model that are within the reach of present observations or of those that will be made possible in the near future by the next generation of space and ground-based telescopes?

These are all questions that we plan to address in future work. Explaining the rich morphology of dwarf galaxies, the wide diversity of their star formation properties, and the scaling laws that link their structural parameters with the properties of their surrounding haloes seems within reach.

\section{Acknowledgements}

We thank the anonymous referee for excellent comments which helped to improve this paper, and Evan Skillman, Dan Weisz and Carme Gallart for valuable feedback on an early draft of this work. We also acknowledge the work of everyone in the EAGLE, APOSTLE, and Auriga collaborations, which have made this work possible. JFN acknowledges the hospitality of the KITP at UC Santa Barbara and of the Aspen Center for Physics. This research was supported in part by the National Science Foundation under Grants No. NSF PHY-1748958 and PHY-1607611, and by the Science and Technology Facilities Council (STFC) consolidated grant ST/P000541/1. CSF acknowledges support by the European Research Council (ERC) through Advanced Investigator grant DMIDAS (GA786910). This work used the DiRAC Data Centric system at Durham University, operated by the Institute for Computational Cosmology on behalf of the STFC DiRAC HPC Facility (\url{www.dirac.ac.uk}). This equipment was funded by BIS National E-infrastructure capital grant ST/K00042X/1, STFC capital grant ST/H008519/1, and STFC DiRAC Operations grant ST/K003267/1 and Durham University. DiRAC is part of the National E-Infrastructure. KO received support from VICI grant 016.130.338 of the Netherlands Foundation for Scientific Research (NWO). FAG acknowledges financial support from CONICYT through the project FONDECYT Regular Nr. 1181264, and funding from the Max Planck Society through a Partner Group grant.


\bibliographystyle{mnras}
\bibliography{bibletter1}


\appendix

\section{Detailed Auriga Results}
\label{appendix_auriga}

Figures~\ref{FigF4F8}-\ref{FigTau90} in the main text compare observations with simulation data from Ap-L1. We include here the same figures, but with data from the Auriga simulation Au-L3. The trends in Au-L3 reproduce well those seen in Ap-L1, and are also seen in the lower-resolution Au-L4. 

\begin{figure*}
\includegraphics[page=7,width=1\linewidth]{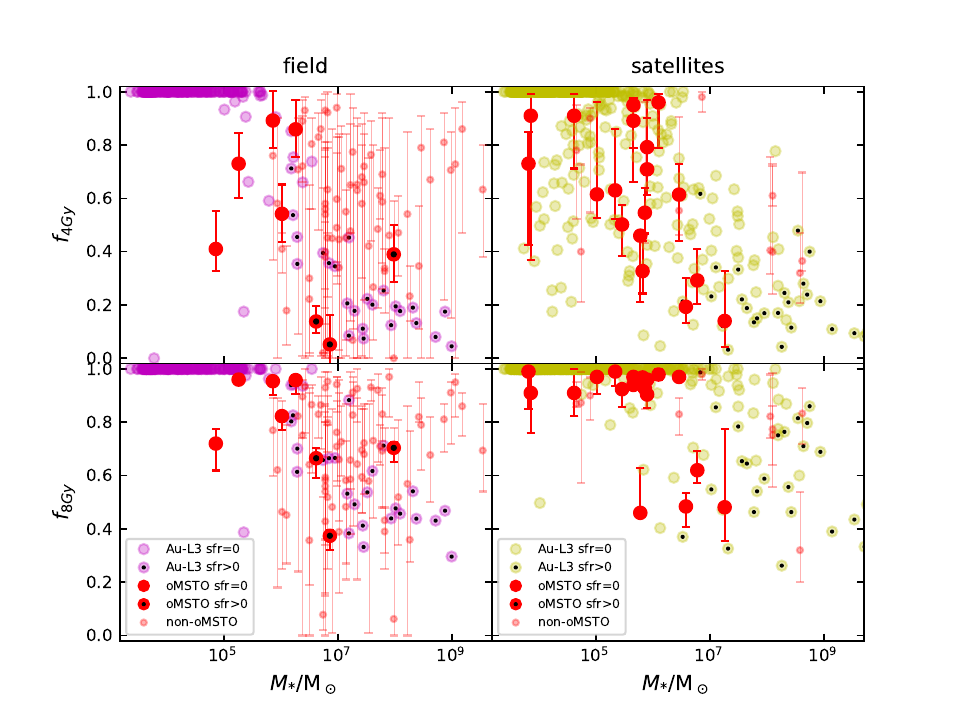}
\caption{As Fig.~\ref{FigF4F8}, but with Au-L3 data: The cumulative fraction of stars formed in the first $4$ ($f_{\rm 4Gy}$) and $8$ ($f_{\rm 8Gy}$) Gyr of cosmic evolution, as a function of stellar mass. Auriga galaxies are shown in magenta (field dwarfs) and yellow (satellites); observed galaxies are in red. Error bars in the latter indicate the $16^{\rm th}$ and $84^{\rm th}$ percentile bounds on the combined statistical and systematic uncertainties, as given in the literature (see Tables~\ref{TabField} and ~\ref{TabSats}). Filled red circles highlight observed galaxies where the photometry reaches the oldest main sequence turnoff.} 
\label{FigAuF4F8} 
\end{figure*}

\begin{figure*}
\includegraphics[page=8, width=1\linewidth, trim = 0cm 5cm 0cm 0cm, clip]{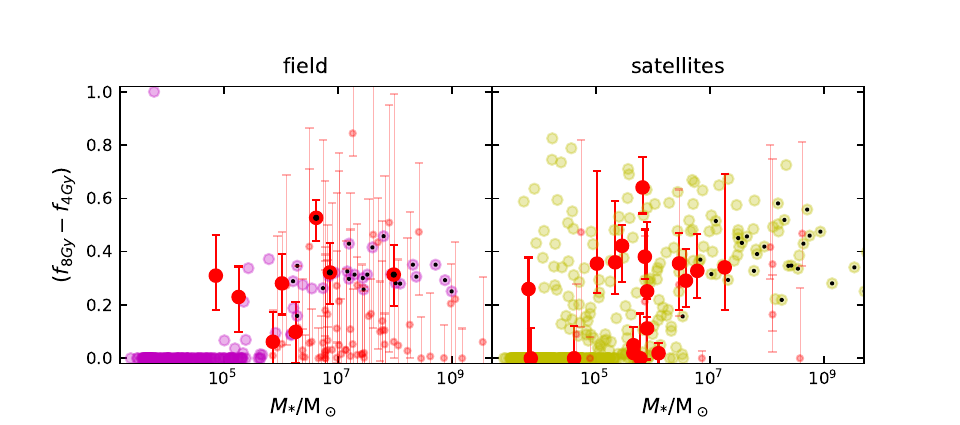}
\caption{As Fig.~\ref{FigAuF4F8}, but for the difference between the fraction of stars formed by the first $4$ and $8$ Gyrs of cosmic evolution.} 
\label{FigAuF4-F8} 
\end{figure*}

\begin{figure*}
\includegraphics[page=9, width=1\linewidth, trim = 0cm 5cm 0cm 0cm, clip]{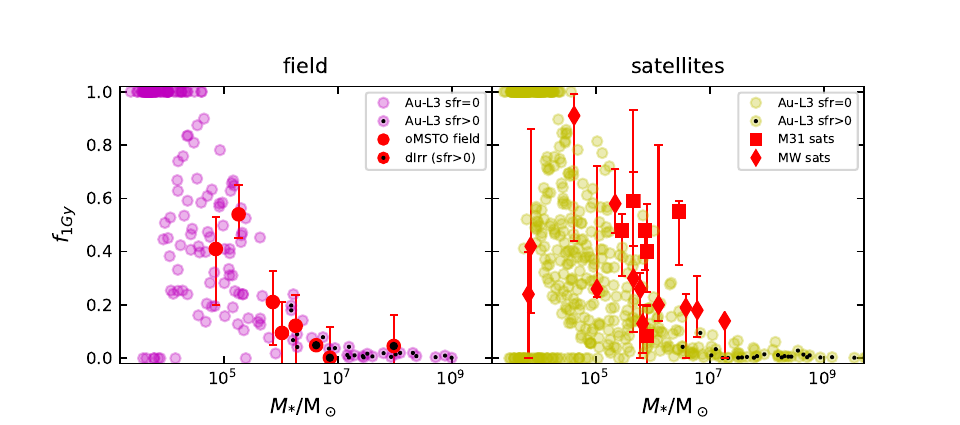} 
\caption{As Fig~\ref{FigMstarf1}, but with Au-L3 data: The fraction of stars formed in the first $\sim 1$ Gy of cosmic evolution, as a function of stellar mass.Error bars indicate the $16^{\rm th}$ and $84^{\rm th}$ percentile bounds on the combined statistical and systematic uncertainties, as published in the literature. Symbols differentiate observed field dwarfs (circles) from  satellites of M31 (squares) and of the Milky Way (diamonds).  Black dots indicate the observed dIrrs Aquarius, IC1613, and LeoA, which are still forming stars at the present day.  Note that Au-L3 field results are a closer match to the observations than the APOSTLE galaxies shown in Fig~\ref{FigAuMstarf1}; Ap-L1 results were dominated by $f_{1 \rm Gy} \sim 0$ at low masses, likely due to the choice of reionization redshift. Like APOSTLE, Au-L3 satellites show a slight but systematic shift toward lower values of $f_{\rm 1Gy}$ or $M_*$. Possible reasons for this include the effects of tidal stripping, or, more likely, inaccuracies related to numerical limitations.  Note that galaxies with $M_{\rm star} < 10^5 M_{\odot}$ are resolved with $\sim 15$ particles or fewer, and are included only for illustration. } \label{FigAuMstarf1}
\end{figure*}

\begin{figure*} \includegraphics[page=10,width=1\linewidth, trim = 0cm 5.cm 0cm 1cm, clip]{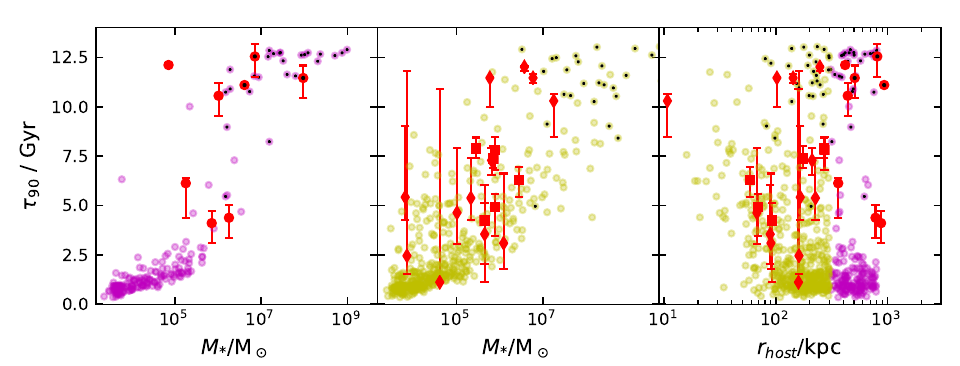} 
\caption{As Fig.~\ref{FigTau90}, but with Au-L3 data: The cosmic time at which galaxies have formed $90\%$ of their stars, $\tau_{90}$, as a function of stellar mass (left and middle panels, showing field and satellite dwarfs, respectively) and as a function of distance from the nearest primary (right-hand panel). } \label{FigAuTau90}
\end{figure*}

\section{Observational Data}
\label{appendix_obsns}

Tables~\ref{TabField} and \ref{TabSats} list the properties of observed field dwarfs and satellites, respectively, used in our analysis. Galaxies are listed alphabetically by name. Values of $f_{1\rm Gy}$ and $\tau_{90}$ are only computed for galaxies that resolve the oMSTO. 

\onecolumn
\begin{ThreePartTable}
\begin{TableNotes}
    \footnotesize
    \item {\bf References}: Stellar masses are derived from B-magnitudes taken from 
     \citet{Karachentsev2013_UpdatedNGC}, assuming a mass-to-light ratio of 1. We take star 
     formation histories from the following references: W11: \citet{Weisz2011_SFHof60DGs}, W14: 
     \citet{Weisz2014_SFHofLGDGs}, C14: \citet{Cole2014_Aquarius}, 
     G15: \citet{Gallart2015_FastSlow}, and S17: \citet{Skillman2017_SFHislands}. 
     Errors indicate the $16^{\rm th}$ and $84^{\rm th}$ percentile bounds on the combined 
     random and systematic errors. SFHs published in \citet{Gallart2015_FastSlow} 
     only quote random errors, whereas the others publish both random and systematic uncertainties; 
     to be consistent in our analysis, we assign galaxies from \citet{Gallart2015_FastSlow} the median 
     error range of the other oMSTO galaxies. These errors are $f_{1Gy}/\rm{Gy}=X^{+0.11}_{-0.16}$; 
     $\tau_{90}/\rm{Gy}=Y^{+0.63}_{-1.03}$.
\end{TableNotes}

\begin{longtable}{{l} *{9}{c} {r}}
    \caption{Data values for the observed \textit{field} galaxies.}\\
    \hline
    \hline
    Gal. Name & BMag & Mstar & $f_{1Gy}$ & $f_{4Gy}$ & $f_{8Gy}$ & $\tau_{90}$ & oMSTO & Ref.\\
    & (mag) & ($M_\odot$) & & & & (Gy) & & \\
    \endfirsthead
    
    {{\bfseries \tablename\ \thetable{} -- (continued)}} \\
    \hline
    \hline
    Gal. Name & BMag & Mstar & $f_{1Gy}$ & $f_{4Gy}$ & $f_{8Gy}$ & $\tau_{90}$ & oMSTO & Ref.\\
    & (mag) & ($M_\odot$) & & & & (Gy) & & \\
    \hline
    \endhead
    
    \insertTableNotes  
    \endlastfoot
  
    \hline 
    \nopagebreak
    A0952+69 & -11.5 & 5.97e+06 & - & $0.4_{-0.0}^{+-0.29}$ & $0.43_{-0.0}^{+-0.06}$ & - & n & W11 \\
AndXXVIII & -7.7 & 1.80e+05 & $0.54_{-0.09}^{+0.11}$ & $0.73_{-0.6}^{+-0.62}$ & $0.96_{-0.96}^{+-0.95}$ & $6.13_{-1.75}^{+0.28}$ & y & S17 \\
Antlia & -9.8 & 1.25e+06 & - & $0.18_{-0.05}^{+0.06}$ & $0.45_{-0.27}^{+-0.11}$ & - & n & W11 \\
Aquarius & -11.1 & 4.13e+06 & $0.05_{-0.02}^{+0.01}$ & $0.14_{-0.1}^{+-0.08}$ & $0.67_{-0.59}^{+-0.63}$ & $11.1_{-0.04}^{+0.02}$ & y & C14 \\
BK3N & -9.6 & 1.04e+06 & - & $0.41_{-0.32}^{+-0.35}$ & $0.46_{-0.25}^{+-0.05}$ & - & n & W11 \\
BK5N & -10.6 & 2.61e+06 & - & $0.93_{-0.0}^{+-0.88}$ & $0.93_{-0.76}^{+-0.87}$ & - & n & W11 \\
Cetus & -10.2 & 1.80e+06 & $0.12_{-0.16}^{+0.11}$ & $0.86_{-0.75}^{+-0.75}$ & $0.96_{-0.91}^{+-0.94}$ & $4.38_{-1.03}^{+0.63}$ & y & G15 \\
DDO113 & -11.5 & 5.97e+06 & - & $0.58_{-0.02}^{+-0.35}$ & $0.59_{-0.36}^{+-0.33}$ & - & n & W11 \\
DDO125 & -14.3 & 7.87e+07 & - & $0.46_{-0.0}^{+-0.02}$ & $0.97_{-0.82}^{+-0.94}$ & - & n & W11 \\
DDO155 & -12.0 & 9.46e+06 & - & $0.6_{-0.36}^{+-0.36}$ & $0.71_{-0.58}^{+-0.54}$ & - & n & W11 \\
DDO165 & -15.1 & 1.64e+08 & - & $0.54_{-0.0}^{+-0.23}$ & $0.57_{-0.0}^{+-0.28}$ & - & n & W11 \\
DDO181 & -13.2 & 2.86e+07 & - & $0.72_{-0.0}^{+-0.52}$ & $0.72_{-0.53}^{+-0.57}$ & - & n & W11 \\
DDO183 & -13.2 & 2.86e+07 & - & $0.66_{-0.22}^{+-0.48}$ & $0.68_{-0.58}^{+-0.51}$ & - & n & W11 \\
DDO187 & -12.4 & 1.37e+07 & - & $0.45_{-0.07}^{+-0.27}$ & $0.48_{-0.27}^{+-0.24}$ & - & n & W11 \\
DDO190 & -14.1 & 6.55e+07 & - & $0.33_{-0.17}^{+0.14}$ & $0.43_{-0.25}^{+-0.08}$ & - & n & W11 \\
DDO44 & -12.1 & 1.04e+07 & - & $0.34_{-0.03}^{+0.16}$ & $0.39_{-0.16}^{+0.13}$ & - & n & W11 \\
DDO53 & -13.4 & 3.44e+07 & - & $0.42_{-0.0}^{+-0.09}$ & $0.58_{-0.01}^{+-0.29}$ & - & n & W11 \\
DDO6 & -12.4 & 1.37e+07 & - & $0.55_{-0.01}^{+-0.26}$ & $0.58_{-0.21}^{+-0.29}$ & - & n & W11 \\
DDO71 & -12.1 & 1.04e+07 & - & $0.45_{-0.0}^{+-0.06}$ & $0.66_{-0.43}^{+-0.36}$ & - & n & W11 \\
DDO78 & -11.5 & 5.97e+06 & - & $0.56_{-0.26}^{+-0.34}$ & $0.58_{-0.34}^{+-0.37}$ & - & n & W11 \\
DDO82 & -14.7 & 1.14e+08 & - & $0.47_{-0.31}^{+-0.33}$ & $0.52_{-0.34}^{+-0.23}$ & - & n & W11 \\
DDO99 & -13.5 & 3.77e+07 & - & $0.76_{-0.44}^{+-0.64}$ & $0.93_{-0.88}^{+-0.9}$ & - & n & W11 \\
ESO269-037 & -12.0 & 9.46e+06 & - & $0.94_{-0.31}^{+-0.89}$ & $0.94_{-0.84}^{+-0.91}$ & - & n & W11 \\
ESO294-010 & -10.9 & 3.44e+06 & - & $0.8_{-0.56}^{+-0.71}$ & $0.87_{-0.61}^{+-0.79}$ & - & n & W11 \\
ESO321-014 & -12.7 & 1.80e+07 & - & $0.77_{-0.11}^{+-0.61}$ & $0.83_{-0.41}^{+-0.7}$ & - & n & W11 \\
ESO325-011 & -14.0 & 5.97e+07 & - & $0.59_{-0.26}^{+-0.25}$ & $0.69_{-0.48}^{+-0.46}$ & - & n & W11 \\
ESO383-087 & -17.0 & 9.46e+08 & - & $0.71_{-0.15}^{+-0.56}$ & $0.91_{-0.72}^{+-0.87}$ & - & n & W11 \\
ESO410-005 & -11.6 & 6.55e+06 & - & $0.63_{-0.47}^{+-0.46}$ & $0.8_{-0.69}^{+-0.7}$ & - & n & W11 \\
ESO540-030 & -11.4 & 5.45e+06 & - & $0.02_{-0.02}^{+0.6}$ & $0.08_{-0.02}^{+0.36}$ & - & n & W11 \\
ESO540-032 & -11.3 & 4.97e+06 & - & $0.86_{-0.01}^{+-0.75}$ & $0.86_{-0.7}^{+-0.76}$ & - & n & W11 \\
F8D1 & -12.6 & 1.64e+07 & - & $0.65_{-0.47}^{+-0.31}$ & $0.66_{-0.55}^{+-0.33}$ & - & n & W11 \\
FM1 & -10.5 & 2.38e+06 & - & $0.89_{-0.66}^{+-0.81}$ & $0.9_{-0.74}^{+-0.83}$ & - & n & W11 \\
    \pagebreak
    HS117 & -11.2 & 4.53e+06 & - & $0.83_{-0.48}^{+-0.74}$ & $0.83_{-0.74}^{+-0.76}$ & - & n & W11 \\
HoI & -14.5 & 9.46e+07 & - & $0.0_{-0.0}^{+0.59}$ & $0.06_{-0.0}^{+0.66}$ & - & n & W11 \\
HoII & -16.7 & 7.18e+08 & - & $0.81_{-0.22}^{+-0.7}$ & $0.81_{-0.69}^{+-0.75}$ & - & n & W11 \\
HoIX & -13.6 & 4.13e+07 & - & $0.27_{-0.07}^{+0.27}$ & $0.73_{-0.44}^{+-0.53}$ & - & n & W11 \\
IC1613 & -14.5 & 9.46e+07 & $0.05_{-0.16}^{+0.11}$ & $0.39_{-0.29}^{+-0.28}$ & $0.7_{-0.65}^{+-0.68}$ & $11.46_{-1.03}^{+0.63}$ & y & G15 \\
IC2574 & -17.5 & 1.50e+09 & - & $0.86_{-0.47}^{+-0.76}$ & $0.86_{-0.75}^{+-0.81}$ & - & n & W11 \\
IC5152 & -15.6 & 2.61e+08 & - & $0.35_{-0.11}^{+-0.12}$ & $0.82_{-0.8}^{+-0.69}$ & - & n & W11 \\
IKN & -11.6 & 6.55e+06 & - & $0.92_{-0.23}^{+-0.84}$ & $0.95_{-0.82}^{+-0.92}$ & - & n & W11 \\
KDG52 & -11.5 & 5.97e+06 & - & $0.93_{-0.19}^{+-0.87}$ & $0.93_{-0.67}^{+-0.87}$ & - & n & W11 \\
KDG61 & -12.9 & 2.17e+07 & - & $0.63_{-0.0}^{+-0.36}$ & $0.64_{-0.21}^{+-0.4}$ & - & n & W11 \\
KDG64 & -12.6 & 1.64e+07 & - & $0.48_{-0.14}^{+-0.07}$ & $0.59_{-0.38}^{+-0.27}$ & - & n & W11 \\
KDG73 & -10.8 & 3.13e+06 & - & $0.3_{-0.01}^{+0.05}$ & $0.36_{-0.0}^{+0.2}$ & - & n & W11 \\
KK077 & -12.0 & 9.46e+06 & - & $0.49_{-0.28}^{+-0.21}$ & $0.76_{-0.56}^{+-0.55}$ & - & n & W11 \\
KKH37 & -11.6 & 6.55e+06 & - & $0.45_{-0.01}^{+-0.11}$ & $0.52_{-0.25}^{+-0.25}$ & - & n & W11 \\
KKH86 & -10.3 & 1.98e+06 & - & $0.71_{-0.09}^{+-0.52}$ & $0.82_{-0.24}^{+-0.7}$ & - & n & W11 \\
KKH98 & -10.8 & 3.13e+06 & - & $0.22_{-0.02}^{+0.12}$ & $0.64_{-0.19}^{+-0.35}$ & - & n & W11 \\
KKR25 & -9.4 & 8.63e+05 & - & $0.58_{-0.0}^{+-0.26}$ & $0.62_{-0.18}^{+-0.35}$ & - & n & W11 \\
KKR3 & -9.2 & 7.18e+05 & - & $0.76_{-0.37}^{+-0.64}$ & $0.77_{-0.6}^{+-0.65}$ & - & n & W11 \\
LeoA & -11.7 & 7.18e+06 & $0.0_{-0.16}^{+0.11}$ & $0.05_{--0.05}^{+0.06}$ & $0.37_{-0.32}^{+-0.35}$ & $12.55_{-1.03}^{+0.63}$ & y & G15 \\
LeoT & -6.7 & 7.18e+04 & $0.41_{-0.21}^{+0.12}$ & $0.41_{-0.33}^{+-0.27}$ & $0.72_{-0.62}^{+-0.67}$ & $12.12_{-0.06}^{+0.12}$ & y & W14 \\
NGC2366 & -16.1 & 4.13e+08 & - & $0.67_{-0.0}^{+-0.48}$ & $0.68_{-0.5}^{+-0.53}$ & - & n & W11 \\
NGC3109 & -15.7 & 2.86e+08 & - & $0.79_{-0.0}^{+-0.67}$ & $0.79_{-0.62}^{+-0.69}$ & - & n & W11 \\
NGC3741 & -13.1 & 2.61e+07 & - & $0.68_{-0.3}^{+-0.48}$ & $0.7_{-0.46}^{+-0.53}$ & - & n & W11 \\
NGC4163 & -13.8 & 4.97e+07 & - & $0.48_{-0.19}^{+-0.33}$ & $0.92_{-0.65}^{+-0.86}$ & - & n & W11 \\
NGC4228 & -17.2 & 1.14e+09 & - & $0.73_{-0.0}^{+-0.52}$ & $0.95_{-0.82}^{+-0.92}$ & - & n & W11 \\
NGC55 & -18.4 & 3.44e+09 & - & $0.63_{-0.38}^{+-0.47}$ & $0.69_{-0.54}^{+-0.52}$ & - & n & W11 \\
NGC6822 & -15.2 & 1.80e+08 & - & $0.23_{-0.14}^{+-0.12}$ & $0.36_{-0.28}^{+0.05}$ & - & n & W14 \\
PegasusdIrr & -11.5 & 5.97e+06 & - & $0.54_{-0.13}^{+-0.29}$ & $0.54_{-0.15}^{+-0.24}$ & - & n & W14 \\
Phoenix & -9.6 & 1.04e+06 & $0.1_{-0.16}^{+0.11}$ & $0.54_{-0.44}^{+-0.43}$ & $0.82_{-0.77}^{+-0.8}$ & $10.56_{-1.03}^{+0.63}$ & y & G15 \\
SagDIG & -11.5 & 5.97e+06 & - & $0.38_{-0.0}^{+-0.0}$ & $0.56_{-0.16}^{+-0.25}$ & - & n & W14 \\
Sc22 & -10.5 & 2.38e+06 & - & $0.72_{-0.01}^{+-0.53}$ & $0.75_{-0.0}^{+-0.56}$ & - & n & W11 \\
SexA & -13.9 & 5.45e+07 & - & $0.61_{-0.52}^{+-0.44}$ & $0.71_{-0.64}^{+-0.63}$ & - & n & W14 \\
SexB & -14.0 & 5.97e+07 & - & $0.69_{-0.36}^{+-0.48}$ & $0.83_{-0.73}^{+-0.76}$ & - & n & W14 \\
Tuc & -9.2 & 7.18e+05 & $0.21_{-0.16}^{+0.11}$ & $0.89_{-0.79}^{+-0.78}$ & $0.95_{-0.9}^{+-0.93}$ & $4.11_{-1.03}^{+0.63}$ & y & G15 \\
UA292 & -11.8 & 7.87e+06 & - & $0.45_{-0.0}^{+-0.04}$ & $0.48_{-0.01}^{+-0.11}$ & - & n & W11 \\
UA438 & -12.9 & 2.17e+07 & - & $0.65_{-0.0}^{+-0.43}$ & $0.93_{-0.8}^{+-0.85}$ & - & n & W11 \\
UGC4483 & -12.7 & 1.80e+07 & - & $0.07_{-0.0}^{+0.51}$ & $0.91_{-0.86}^{+-0.85}$ & - & n & W11 \\
UGC8508 & -13.1 & 2.61e+07 & - & $0.58_{-0.01}^{+-0.36}$ & $0.59_{-0.46}^{+-0.34}$ & - & n & W11 \\
UGC8833 & -12.2 & 1.14e+07 & - & $0.71_{-0.0}^{+-0.55}$ & $0.73_{-0.6}^{+-0.59}$ & - & n & W11 \\
WLM & -14.1 & 6.55e+07 & - & $0.34_{-0.24}^{+-0.27}$ & $0.39_{-0.31}^{+-0.3}$ & - & n & W14 \\
    \hline
    \label{TabField}
    
\end{longtable}
\end{ThreePartTable}

\newpage
\onecolumn
\begin{ThreePartTable}
\begin{TableNotes}
    \footnotesize
    \item {\bf References}: Stellar masses are derived from B-magnitudes taken from
    \citet{Karachentsev2013_UpdatedNGC}, assuming a mass-to-light ratio of 1. We take star 
    formation histories from the following references: W11: \citet{Weisz2011_SFHof60DGs}, W14: 
    \citet{Weisz2014_SFHofLGDGs}, G15: \citet{Gallart2015_FastSlow}, and S17: 
    \citet{Skillman2017_SFHislands}. Errors indicate the $16^{\rm th}$ and $84^{\rm th}$ percentile 
     bounds on the combined random and systematic errors.
\end{TableNotes}

\begin{longtable}{{l} *{9}{c} {r}}
    \caption{Data values for the observed \textit{satellite} galaxies.}\\
    \hline
    \hline
    Gal. Name & BMag & Mstar & $f_{1Gy}$ & $f_{4Gy}$ & $f_{8Gy}$ & $\tau_{90}$ & oMSTO & Ref.\\
    & (mag) & ($M_\odot$) & & & & (Gy) & & \\
    \endfirsthead
    
    {{\bfseries \tablename\ \thetable{} -- (continued)}} \\
    \hline
    \hline
    Gal. Name & BMag & Mstar & $f_{1Gy}$ & $f_{4Gy}$ & $f_{8Gy}$ & $\tau_{90}$ & oMSTO & Ref.\\
    & (mag) & ($M_\odot$) & & & & (Gy) & & \\
    \hline
    \endhead
    
    \insertTableNotes  
    \endlastfoot
  
    \hline
    AndI & -10.7 & 2.86e+06 & $0.55_{-0.2}^{+0.04}$ & $0.61_{-0.44}^{+-0.5}$ & $0.97_{-0.96}^{+-0.96}$ & $6.29_{-0.84}^{+0.67}$ & y & S17 \\
AndII & -9.2 & 7.18e+05 & $0.48_{-0.15}^{+0.02}$ & $0.55_{-0.47}^{+-0.45}$ & $0.93_{-0.9}^{+-0.89}$ & $7.39_{-0.51}^{+0.6}$ & y & S17 \\
AndIII & -9.3 & 7.87e+05 & $0.4_{-0.15}^{+0.18}$ & $0.71_{-0.61}^{+-0.45}$ & $0.96_{-0.96}^{+-0.95}$ & $4.93_{-1.47}^{+0.67}$ & y & S17 \\
AndV & -9.2 & 7.18e+05 & - & $0.72_{-0.43}^{+-0.47}$ & $0.93_{-0.85}^{+-0.91}$ & - & n & W14 \\
AndVI & -10.7 & 2.86e+06 & - & $0.55_{-0.46}^{+-0.2}$ & $0.83_{-0.75}^{+-0.77}$ & - & n & W14 \\
AndVII & -11.7 & 7.18e+06 & - & $0.98_{-0.92}^{+-0.96}$ & $0.98_{-0.96}^{+-0.96}$ & - & n & W14 \\
AndXI & -6.2 & 4.53e+04 & - & $0.78_{-0.52}^{+-0.61}$ & $0.87_{-0.86}^{+-0.79}$ & - & n & W14 \\
AndXII & -6.4 & 5.44e+04 & - & $0.4_{-0.21}^{+-0.07}$ & $0.87_{-0.57}^{+-0.76}$ & - & n & W14 \\
AndXIII & -6.8 & 7.87e+04 & - & $0.9_{-0.55}^{+-0.86}$ & $0.9_{-0.81}^{+-0.83}$ & - & n & W14 \\
AndXV & -8.7 & 4.53e+05 & $0.59_{-0.17}^{+0.34}$ & $0.89_{-0.79}^{+-0.82}$ & $0.94_{-0.94}^{+-0.93}$ & $4.24_{-3.13}^{+0.87}$ & y & S17 \\
AndXVI & -8.2 & 2.86e+05 & $0.48_{-0.17}^{+0.06}$ & $0.5_{-0.39}^{+-0.43}$ & $0.92_{-0.86}^{+-0.9}$ & $7.88_{-0.49}^{+0.56}$ & y & S17 \\
CanVenI & -7.9 & 2.17e+05 & $0.58_{-0.11}^{+0.13}$ & $0.63_{-0.52}^{+-0.4}$ & $0.99_{-0.94}^{+-0.98}$ & $5.38_{-1.13}^{+2.01}$ & y & W14 \\
CanVenII & -4.1 & 6.55e+03 & $0.24_{-0.24}^{+0.16}$ & $0.73_{-0.42}^{+-0.61}$ & $0.99_{-0.85}^{+-0.99}$ & $5.42_{-1.15}^{+3.59}$ & y & W14 \\
Car & -9.0 & 5.97e+05 & $0.26_{-0.26}^{+0.06}$ & $0.46_{-0.21}^{+-0.44}$ & $0.46_{-0.46}^{+-0.29}$ & $11.46_{-1.49}^{+0.07}$ & y & W14 \\
Draco & -8.7 & 4.53e+05 & $0.3_{-0.2}^{+0.4}$ & $0.95_{-0.66}^{+-0.92}$ & $0.97_{-0.95}^{+-0.96}$ & $3.55_{-1.52}^{+2.5}$ & y & W14 \\
For & -11.5 & 5.97e+06 & $0.18_{-0.1}^{+0.13}$ & $0.29_{-0.2}^{+-0.17}$ & $0.62_{-0.57}^{+-0.55}$ & $11.46_{-0.27}^{+0.2}$ & y & W14 \\
Her & -6.1 & 4.13e+04 & $0.91_{-0.47}^{+0.08}$ & $0.91_{-0.71}^{+-0.83}$ & $0.91_{-0.82}^{+-0.82}$ & $1.11_{-0.0}^{+9.77}$ & y & W14 \\
IC10 & -16.0 & 3.77e+08 & - & $0.32_{-0.21}^{+-0.17}$ & $0.32_{-0.2}^{+-0.1}$ & - & n & W14 \\
LGS3 & -9.3 & 7.87e+05 & $0.08_{-0.16}^{+0.11}$ & $0.79_{-0.69}^{+-0.68}$ & $0.9_{-0.85}^{+-0.88}$ & $7.82_{-1.03}^{+0.63}$ & y & G15 \\
LeoI & -11.0 & 3.77e+06 & $0.19_{-0.19}^{+0.05}$ & $0.19_{-0.13}^{+-0.09}$ & $0.48_{-0.41}^{+-0.43}$ & $12.02_{-0.2}^{+0.06}$ & y & W14 \\
LeoII & -9.1 & 6.55e+05 & $0.13_{-0.11}^{+0.07}$ & $0.33_{-0.24}^{+-0.22}$ & $0.97_{-0.92}^{+-0.96}$ & $7.29_{-0.75}^{+0.6}$ & y & W14 \\
LeoIV & -4.2 & 7.18e+03 & $0.42_{-0.25}^{+0.44}$ & $0.91_{-0.37}^{+-0.83}$ & $0.91_{-0.76}^{+-0.83}$ & $2.45_{-0.92}^{+9.36}$ & y & W14 \\
M32 & -14.8 & 1.25e+08 & - & $0.61_{-0.57}^{+-0.49}$ & $0.77_{-0.64}^{+-0.69}$ & - & n & W14 \\
NGC147 & -14.8 & 1.25e+08 & - & $0.4_{-0.24}^{+-0.08}$ & $0.75_{-0.56}^{+-0.52}$ & - & n & W14 \\
NGC185 & -14.7 & 1.14e+08 & - & $0.41_{-0.32}^{+-0.05}$ & $0.82_{-0.74}^{+-0.66}$ & - & n & W14 \\
NGC205 & -16.1 & 4.13e+08 & - & $0.36_{-0.33}^{+-0.03}$ & $0.83_{-0.72}^{+-0.74}$ & - & n & W14 \\
SagdSph & -12.7 & 1.80e+07 & $0.14_{-0.14}^{+0.0}$ & $0.14_{-0.04}^{+0.05}$ & $0.48_{-0.35}^{+-0.19}$ & $10.3_{-1.82}^{+0.33}$ & y & W14 \\
Sculptor & -9.8 & 1.25e+06 & $0.2_{-0.06}^{+0.6}$ & $0.96_{-0.79}^{+-0.93}$ & $0.98_{-0.96}^{+-0.96}$ & $3.09_{-1.29}^{+3.53}$ & y & W14 \\
UMi & -7.1 & 1.04e+05 & $0.26_{-0.03}^{+0.46}$ & $0.61_{-0.53}^{+-0.27}$ & $0.97_{-0.91}^{+-0.95}$ & $4.63_{-1.6}^{+3.27}$ & y & W14 \\
    \hline
    \label{TabSats}
    
\end{longtable}
\end{ThreePartTable}


\bsp
\label{lastpage}
\end{document}